\documentstyle[aps,titlepage,12pt]{revtex}

\begin{document} 
\title{Wormholes, Classical Limit and Dynamical Vacuum in Quantum Cosmology}
\author{Nivaldo A. Lemos and Fl\'avio G. Alvarenga\\
{\small{Departamento de F\'{\i}sica}}\\
{\small{Universidade Federal Fluminense}}\\
{\small{Av. Litor\^anea s/n, Boa Viagem - CEP 24210-340}}\\
{\small\hspace{0.5cm}{Niter\'{o}i - Rio de Janeiro}}\\
{\small{Brazil}}}
\maketitle

\begin{abstract}
First a Friedmann-Robertson-Walker (FRW) universe  filled with dust and a  conformally invariant scalar field is quantized. For  the   closed model we find a discrete set of wormhole quantum states. In the case of  flat spacelike sections we find states with classical behaviour at small values of the scale factor and quantum  behaviour for large  values of the scale factor. Next we study a FRW model with a conformally invariant scalar field and a nonvanishing cosmological constant  dynamically introduced by regarding the vacuum  as a perfect fluid with equation of state $p = - \rho$. The ensuing Wheeler-DeWitt equation turns out to be a bona fide Schr\"{o}dinger equation, and we find that there are realizable states with a definite value of the cosmological constant. Once again we find finite-norm solutions to the  Wheeler-DeWitt equation with definite values of the cosmological constant that represent wormholes, suggesting  that 
in quantum cosmological models with a  simple matter content wormhole states are a common occurrence.

\vskip 1.5cm

\end{abstract}

\newpage

\section{Introduction}

Classical wormholes are solutions to the Euclidean Einstein equations consisting of two asymptotically flat regions connected by a throat. However, classical wormholes may only exist for exotic matter content, as for example axionic matter \cite{Strominger}, such as to cause the Ricci tensor to have negative eigenvalues, which prevent them from playing any fundamental role in physical processes. Quantum wormholes have been defined as solutions to the Wheeler-DeWitt equation with suitable boundary conditions. The wave function should be regular when the three-geometry collapses to zero and be exponentially damped for large three-geometries \cite{Hawking}. As opposed to the classical situation, in quantum cosmology wormhole states  are not restricted to  very special kinds of matter fields, and in fact they exist in models with a very simple matter content.

We first study the quantum theory of a Friedmann-Robertson-Walker  universe filled with dust and a  conformally invariant scalar field.  The dust provides a suitable clock, and a time variable can be naturally introduced. The Wheeler-DeWitt equation becomes  a genuine Schr\"{o}dinger equation whose stationary states are investigated. For a closed universe we find a basis of finite-norm eigenfunctions with all the properties of wormhole states. In the spatially flat case we find solutions to the Wheeler-DeWitt equation which display a quantum behaviour for large values of the scale factor but behave classically for small values of the scale 
factor. This unintuitive behaviour is corroborated by a Bohmian analysis,
and provides evidence that the issue of the classical limit in quantum cosmology seems more delicate than one would naively expect.
 
In modern cosmology, especially since the advent of the inflationary model \cite{Guth}, the terms {\it vacuum energy} and {\it cosmological constant} have become almost synonymous \cite{Kolb}. Recently, we have studied \cite{Lemos2}   quantum aspects of de Sitter's cosmological model by treating the vacuum as a dynamical entity. In such a treatment, the cosmological constant is not be postulated from the start, but emerges from the dynamical degrees of freedom of the vacuum. This is achieved by regarding the vacuum as a perfect fluid with equation of state $p=-\rho$. This approach leads to interesting consequences  in inflationary cosmology \cite{Lima}, and has several attractive features in the context of quantum cosmology. In particular, the assignation of dynamical degrees of freedom to the vacuum  makes room for the natural introduction of a time variable.

In this paper we adopt  the point of view described above in order to  add a nonvanishing cosmological constant to a FRW  universe  filled with   a conformally invariant scalar field. Schutz's canonical formalism \cite{Schutz}, which  describes a relativistic fluid interacting with the gravitational field, is employed, since it has the feature of ascribing dynamical degrees of freedom to the fluid. As we have shown before \cite{Lemos2}, Schutz's action principle  is successful even in the case of the vacuum in the sense that  the cosmological constant appears dynamically as a manifestation of the  degrees of freedom of the fluid that plays the role of the vacuum. The quantum  properties of the model are investigated on the basis of the associated Wheeler-DeWitt equation. Because the super-Hamiltonian constraint is linear in one of the momenta, the Wheeler-DeWitt equation can be reduced to a bona fide Schr\"odinger equation. In contrast to the case of the de Sitter model \cite{!
Lemo
s2}, there exist finite-norm solutions to the Wheeler-DeWitt  equation with definite values of the  cosmological constant. We find an infinite set  of quantum wormhole states for certain discrete positive  values of
the cosmological constant.

This paper is organized as follows. In Section 2 a Hamiltonian treatment is given to a FRW cosmological model with dust and a conformally invariant scalar field as material sources. Then, the model is quantized and, in the case of closed universes, a set of finite-norm solutions to the Wheeler-DeWitt equation are found which possess all the properties usually ascribed to wormhole quantum states. In the spatially flat case solutions are found with a classical behaviour for small scale factor and a quantum behaviour for large scale factor. A Bohmian analysis  confirms this peculiar behaviour. In Section 3 a cosmological term is dynamically added to a FRW universe filled with a conformally invariant scalar field.
In the case of closed universes it is shown that, for certain discrete positive values of the cosmological constant, there exist finite-norm solutions to the Wheeler-DeWitt equation that represent wormholes. Section 4 is dedicated to  final comments.

\section{FRW Model With Conformal Scalar Field and Dust}

The metric of a homogeneous and isotropic cosmological model
can be written in the Friedmann-Robertson-Walker form (we take $c=1$)
\begin{equation}
\label{metric}
ds^2=-N(t)^2a(t)^2dt^2 +  a(t)^2\sigma_{ij}dx^{i}dx^{j},
\end{equation}

\noindent 
where ${\sigma}_{ij}$ denotes the   metric for a 3-space of constant curvature $k= +1, 0$ or $-1$, corresponding to spherical, flat or hyperbolic spacelike sections, respectively. The lapse function has been conveniently parametrized as $N(t)a(t)$.          

In units such that $c=16\pi G=1$, the pure gravitational action is
\begin{equation}
S_{g}=\int_{M}^{}d^4 x \sqrt{-g}\hspace{0.2cm}R  + 2\int_{\partial M}^{} d^3 x \sqrt{h}\hspace{0.2cm} K\hspace{0.3cm},
\end{equation}

\noindent 
where $K$ is the trace of the extrinsic curvature $K_{ij}$ of the boundary $\partial M$ of the space-time manifold $M$.

The matter content will be taken to be a perfect fluid plus  a scalar field conformally coupled to gravity. The action associated with the sources of gravity  is
\begin{displaymath}
\label{af}
S_{m}=\int_{M}^{}d^4 x \sqrt{-g}\hspace{0.3cm}p 
\end{displaymath}
\begin{equation}
- \frac{1}{2}\int_{M}^{}d^4 x \sqrt{-g}\hspace{0.3cm}\bigg(\partial_{\mu}\phi\partial^{\mu}\phi + \frac{1}{6}R\phi^2\bigg) -\frac{1}{12}\int_{\partial M}^{}d^3 x \sqrt{h}\hspace{0.3cm}K\phi^2\hspace{0.3cm}.
\end{equation}

\noindent 
where Schutz's canonical formulation of the dynamics of a relativistic fluid in interaction with the gravitational field is  being employed  \cite{Schutz}. This formalism makes use of a representation  for the four-velocity of the fluid in terms of the five velocity potentials
$\epsilon, \alpha, \beta, \theta$ and $S$ in the form 
\begin{equation}
U_{\nu} = \frac{1}{\mu}\hspace{0.2cm}({\epsilon ,}_{\nu} + \alpha {\beta ,}_{\nu} + \theta {S ,}_{\nu})\hspace{0.5cm},
\end{equation}

\noindent 
where $\mu$ is the specific enthalpy, which is expressed in terms of the velocity potentials by means of the normalization condition $U^{\nu}U_{\nu}=-1$. In the FRW model the potentials $\alpha$ and $\beta$ are zero, since they describe vortex motions. The potential $S$ is the specific entropy, while $\theta$ and $\epsilon$  have no obvious physical interpretation.

Compatibility with the homogeneous spacetime metric is guaranteed 
by taking the scalar field and all of the velocity potentials of the fluid as functions of $t$ only. We shall take $p = (\gamma - 1)\, \rho$ as  equation of state for the fluid, where $\gamma$ is a constant and $\rho$ is the fluid's energy density (we shall eventually put $\gamma = 1$).
Performing an ADM reduction described in detail in \cite{Rubakov,Lemos3}, and setting $\gamma =1$ for dust, we can write the full action $(S_{g} + S_{m})$ as \cite{Rubakov,Lemos3,Peleg}
\begin{equation}
S=\int_{}^{} dt (p_{a}\dot{a} + p_{\Phi}\dot{\Phi} +  p_{\epsilon}\dot{\epsilon} - N\cal{H})\hspace{0.3cm},
\end{equation}

\noindent 
with the super-Hamiltonian $\cal{H}$  given by
\begin{equation}
\label{superhamiltonian}
{\cal{H}}= - \bigg(\frac{{p_{a}}^2}{24} + 6ka^2\bigg) + \bigg(\frac{p_{\Phi}^2}{24} + 6k{\Phi}^2\bigg) + a\, p_{\epsilon}\hspace{0.3cm},
\end{equation}

\noindent 
where we have introduced  the  momenta conjugate to the canonical variables corresponding to the degrees of freedom ascribed to the model,
and  redefined the scalar field by means of $\Phi=\pi a \phi$.

The classical equations of motion are
\begin{equation}
\label{ec1}
\dot{a} = - \frac{N}{12} p_{a}\hspace{0.3cm},\hspace{0.3cm}{\dot{p}}_{a} = 12k N a - N p_{\epsilon}\hspace{0.5cm},
\end{equation}

\begin{equation}
\label{ec2}
{\dot{\Phi}} = \frac{N}{12} p_{\Phi}\hspace{0.3cm},\hspace{0.3cm}{\dot{p}}_{\Phi} = - 12kN\Phi\hspace{0.5cm},
\end{equation}

\begin{equation}
\label{ec3}
\dot{\epsilon} = N a\hspace{0.3cm},\hspace{0.3cm}{\dot{p}}_{\epsilon} = 0\hspace{0.2cm}\Longrightarrow\hspace{0.2cm}p_{\epsilon} = \mbox{const.}\hspace{0.5cm},
\end{equation}

\noindent 
supplemented by the super-Hamiltonian constraint
\begin{equation}
\label{ec4}
{\cal H} = - \bigg(\frac{{p_{a}}^2}{24} + 6ka^2\bigg) + \bigg(\frac{p_{\Phi}^2}{24} + 6k{\Phi}^2\bigg) + a\, p_{\epsilon} = 0 \hspace{0.3cm}.
\end{equation}

In order to solve these equations let us choose the conformal gauge $N=1$. It follows from eqs. (\ref{ec1}) and (\ref{ec2}) that
\begin{equation}
\label{ec5}
\dot{a} = - \frac{1}{12} p_{a}\hspace{0.3cm},\hspace{0.3cm}{\dot{p}}_{a} = 12ka -  p_{\epsilon}\hspace{0.5cm},
\end{equation}

\begin{equation}
\label{ec6}
\ddot{\Phi} + k\,\,\, \Phi = 0 \hspace{0.5cm}.
\end{equation}

First  we  consider the case $k=0$. The equations of motion for $a(t)$ and $\chi(t)$ admit the  simple  solutions
\begin{equation}
\label{escala}
a(t) = A + B\,\,\, t + \frac{p_{\epsilon}}{24}\,\,\,t^2\hspace{0.5cm},
\end{equation}

\begin{equation}
\Phi(t) = C +  D\,\,\, t\hspace{0.5cm},
\end{equation}

\noindent 
with the constraint 
\begin{equation}
\label{vinculo1}
p_{\epsilon}= 6 (B^2 - D^2)/{A}\hspace{0.5cm}.
\end{equation}

For the case $k=+1$ the classical solutions are
\begin{equation}
a(t) = A_{1}\,\,\, \sin (t + \delta_{1}) + \frac{p_{\epsilon}}{12}\hspace{0.5cm},
\end{equation}

\begin{equation}
\Phi(t) = A_{2}\,\,\, \sin (t + \delta_{2})\hspace{0.5cm},
\end{equation}

\noindent 
where 
\begin{equation}
\label{vinculo2}
p_{\epsilon}= 12\sqrt{{A_{1}}^2 - {A_{2}}^2}\hspace{0.5cm}.
\end{equation}

Finally, we find the following solutions for the case $k=-1$:
\begin{equation}
a(t) = A_{3}\,\,\, \sinh (t + \delta_{3}) + \frac{p_{\epsilon}}{12}\hspace{0.5cm},
\end{equation}

\begin{equation}
\Phi(t) = A_{4}\,\,\, \sinh (t + \delta_{4})\hspace{0.5cm},
\end{equation}

\noindent 
with 
\begin{equation}
\label{vinculo3}
p_{\epsilon}= 12\sqrt{{A_{4}}^2 - {A_{3}}^2}\hspace{0.5cm}.
\end{equation}

\section{Quantization}

The quantization of the model in the Wheeler-DeWitt scheme consists in setting
\begin{equation}
p_{a}\rightarrow -i\frac{\partial}{\partial a}\hspace{0.5cm},\hspace{0.5cm}p_{\Phi}\rightarrow -i\frac{\partial}{\partial \Phi}\hspace{0.5cm},\hspace{0.5cm}p_{\epsilon}\rightarrow -i\frac{\partial}{\partial \epsilon}\hspace{0.5cm}
\end{equation}

\noindent 
to form the operator $\hat {\cal{H}}$, and  imposing the Wheeler-DeWitt equation 
\begin{equation}
{\hat{{\cal{H}}}}\Psi=0
\end{equation}

\noindent 
on the wave function of the Universe $\Psi$. In our present case this equation takes the form of a Schr\"odinger-like equation 
\begin{displaymath}
\bigg[\bigg(\frac{1}{24}\frac{{\partial}^2}{\partial a^2} - 6{k}{a^2}\bigg) - \bigg(\frac{1}{24}\frac{{\partial}^2}{\partial {\Phi}^2} - 6k{\Phi}^2 \bigg)\bigg]\Psi(a,\Phi,\epsilon )
\end{displaymath}
\begin{equation}
\label{es1}
 = i\,\,a\,\,\frac{\partial}{\partial \epsilon} \Psi(a,\Phi,\epsilon )\hspace{0.3cm}
\end{equation}

\noindent 
with $\epsilon$ playing the role of time.

The rescaling of variables
\begin{equation}
a=\frac{R}{\sqrt{12}}\hspace{0.5cm},\hspace{0.5cm}\Phi=\frac{\chi}{\sqrt{12}}\hspace{0.5cm}\mbox{and}\hspace{0.5cm} {\epsilon}=\frac{t}{\sqrt{12}}\hspace{0.6cm},
\end{equation}

\noindent 
takes  equation (\ref{es1}) to the form
\begin{displaymath}
\bigg[\bigg(\frac{1}{2}\frac{{\partial}^2}{\partial R^2} - \frac{1}{2}k R^2\bigg) - \bigg(\frac{1}{2}\frac{{\partial}^2}{\partial {\chi}^2} - \frac{1}{2}k{\chi}^2\bigg)\bigg]\Psi(R,\chi,t)
\end{displaymath}
\begin{equation}
\label{es2}
 = R\hspace{0.2cm}i\hspace{0.15cm}\frac{\partial}{\partial t} \Psi(R,\chi,t)\hspace{0.3cm}.
\end{equation}

We try to solve (\ref{es2}) by separation of variables  writing  $\Psi(R,\chi,t)=\psi_{R}(R)\psi_{\chi}(\chi)e^{-iEt}$. This gives rise to the following ordinary differential equations:
\begin{equation}
\label{es3}
\frac{1}{\Psi_{\chi}(\chi )}\bigg( -\frac{1}{2}\frac{d^2} {d{\chi}^2} +\frac{1}{2}k{\chi}^2\bigg)\Psi_{\chi}(\chi )  = \lambda \hspace{0.6cm} ,
\end{equation}

\begin{equation}
\label{es4}
\frac{1}{\Psi_{R}(R)}\bigg(\frac{1}{2} \frac{d^2}{dR^2}  
-\frac{1}{2}kR^2\bigg)\Psi_{R}(R) - ER = -\lambda \hspace{0.6cm} .
\end{equation}

\noindent 
where $\lambda$ is a separation constant.

We first consider the case $k=1$. Then equations (\ref{es3}) and (\ref{es4}) reduce  to 
\begin{equation}
\label{osc1}
\bigg(-\frac{1}{2}\frac{d^2}{d{\chi}^2} + \frac{1}{2}{\chi}^2  \bigg)\Psi_{\chi}(\chi) = \lambda \, \Psi_{\chi}(\chi)
\end{equation}

\noindent 
and
\begin{equation}
\label{osc2}
\frac{d^2\Psi_{R}(R)}{dR^2} + (2\lambda-2ER-R^2)\Psi_{R}(R)=0 \hspace{0.6cm}
\end{equation}

The redefinitions
\begin{equation}
\label{redef}
x=R+E\hspace{0.6cm} , \hspace{0.6cm} \lambda^{\prime}= 2\lambda + E^2
\end{equation}

\noindent 
cast Eq.(\ref{osc2}) into the form
\begin{equation}
\label{osc3}
\frac{d^{2}\Psi_{R}(x)}{dx^2} + (\lambda^{\prime}-x^2)\Psi_{R}(x)=0\hspace{0.6cm}. 
\end{equation}

\noindent 
A comparison of Eqs. (\ref{osc1}) and (\ref{osc3}) with the standard Schr\"{o}dinger equation for a harmonic oscillator gives the allowed values of $\, \lambda\,$ and $\, \lambda^{\prime}\,$:
\begin{equation}
\lambda = n+\frac{1}{2} \,\,\,\, ,\,\,\, n=0,1,2,\ldots \,\,\,\,\,\,\,\, ; \,\,\,\,\,\,\,\, \lambda^{\prime}= 2(m+\frac{1}{2})\,\,\,\, ,\,\,\, m=0,1,2,\ldots
\label{valorlambda}
\end{equation}

\noindent 
In virtue of Eq. (\ref{redef}), the possible values of $\, E\,$ are
\begin{equation}
\label{autovalores1}
E = \pm E_{mn}
\end{equation} 

\noindent 
with
\begin{equation}
\label{autovalores2}
 E_{mn} =  \sqrt{2(m-n)}\hspace{0.5cm}, \hspace{0.5cm} m,n= 0,1,2,\ldots \,\, (m\geq n)\hspace{0.5cm}.
\end{equation} 

A set of normalizable solutions to equation (\ref{es2}) is therefore given by 
\begin{equation}
\label{worm}
 \Psi^{\pm}_{nm}(R,\chi ,t) =
 \varphi_{n}(R \pm E_{mn})
\hspace{0.2cm}\varphi_{m}(\chi)\hspace{0.2cm}e^{\mp\, i\, E_{mn}\, t}
\end{equation}

\noindent 
with $\varphi_{n}(x)$ the harmonic-oscillator wave functions
\begin{equation}
\varphi_{n}(x)=\frac{(-1)^{n}}{\sqrt{2^{n} n! \sqrt{\pi}}}\hspace{0.2cm}e^{x^2/2}\hspace{0.2cm}\frac{d^n}{d x^n} (e^{-x^2})\hspace{0.3cm},\hspace{0.3cm}n=0,1,2,...
\end{equation}

\noindent 
If the scale factor $\, R\,$ is allowed to run over the whole real line as in \cite{Garay2}, all of the  wave functions (\ref{worm}) are acceptable. Furthermore, they can be interpreted as  wormhole states  \cite{Hawking} since they are exponentially damped for large three-geometries ($R\rightarrow \infty$), are regular over the whole configuration space $(R,\phi )= R\!\!\!\!I^{\, 2}$, and have a well-defined limit as $R$ tends to zero, that is, no singularities are developed when the three-geometry degenerates. Here these wormhole wave functions are square-integrable, the integration extending over the entire configuration space. This is quite reasonable in the Schr\"{o}dinger picture of quantum mechanics since the ``time" $t$ is an external  parameter with respect to the configuration space. The situation here should be contrasted with the requirement of square-integrability over the whole configuration space when time is an {\it internal} parameter \cite{Garay1,Garay2,Mena}!
, wh
ose physical meaning is far from clear, and deviates from the well-established canons of quantum mechanics.

At this point it is fitting to address an important issue. The arguments in \cite{Garay2} are based on the theory of hyperbolic partial differential equations and do not seem to apply to our case, so that here the configuration space should be taken to be $\{ R>0,\phi \in R\!\!\!\!I\,\} $. In order for Eq.(\ref{es2}) to be a genuine Schr\"odinger equation
\begin{equation}
\label{Schrodinger}
 i\frac{\partial \Psi}{\partial t} = {\hat H}\, \Psi
\end{equation}

\noindent 
with $t$ a legitimate time, the Hamiltonian operator
\begin{equation}
\label{operhamilt}
{\hat H}=\frac{1}{2R}\frac{{\partial}^2}{\partial R^2} - \frac{1}{2}k R - \frac{1}{R}\bigg(\frac{1}{2}\frac{{\partial}^2}{\partial {\chi}^2} - \frac{1}{2}k{\chi}^2\bigg)
\end{equation}

\noindent 
must be self-adjoint. Following the similar discussion in \cite{Lemos2}, this Hamiltonian operator is self-adjoint as long as the inner-product is chosen as
\begin{equation}
\label{inner}
(\Psi ,\Phi ) = \int_{-\infty}^{\infty}d\chi \int_0^{\infty} dR \, R\, \Psi^*(R,\chi ,t)\, \Phi (R,\chi ,t)
\end{equation}

\vspace{0.3cm}

\noindent
and the domain of $\,{\hat H}\,$ is restricted to those functions such that
\begin{equation}
\label{boundary1}
\Psi (0,\chi ,t)=0
\end{equation}

\noindent 
or 
\begin{equation}
\label{boundary2}
\Psi^{\prime}(0,\chi ,t)=0
\,\,\, ,
\end{equation}

\noindent 
the prime denoting partial derivative with respect to $R$. For the sake of simplicity, we refrain from considering the  general case $\Psi^{\prime}(0,\chi ,t)=\alpha\,\Psi (0,\chi ,t)$ with $\alpha\in (-\infty , \infty ]$. Now a simple analysis shows that no superposition of the states (\ref{worm}) can satisfy the boundary conditions (\ref{boundary1}) or (\ref{boundary2}) for all time  except if $\, m=n\,$, so that $\, E=0\,$. Thus our set of allowed wormhole states reduce 
to those originally found by Hawking and Page \cite{Hawking} with two differences. The first one is that only odd or even values of $\, m=n\,$ are allowed in (\ref{worm}) according to whether the boundary condition (\ref{boundary1}) or (\ref{boundary2}) is selected. The second one is that here the requirement of being normalizable in the inner product (\ref{inner}) is sensible, since the integration extends over a  configuration space whith respect to which time is an external parameter. 
 
We now investigate the case  $k=0$. We have the  uncoupled ordinary differential equations
\begin{equation}
\label{es5}
\frac{d^2 \Psi_{\chi}(\chi)}{d{\chi}^2} + {\mu}^2\,\, \Psi_{\chi}(\chi)  = 0\hspace{0.5cm} ,
\end{equation}

\begin{equation}
\label{es6}
\frac{d^2 \Psi_{R}(R)}{dR^2} - (2ER - {\mu}^2)\,\,  
\Psi_{R}(R) = 0 \hspace{0.6cm} ,
\end{equation}

\noindent 
where we have written $\mu^2$ for the separation constant.
For Eq. (\ref{es5}) there is a free-particle solution
\begin{equation} 
\Psi_{\chi}(\chi) = e^{i\,\mu\,\chi}\hspace{0.5cm}.
\end{equation}

\noindent 
In order to solve Eq. (\ref{es6}), let us  replace the variable $R$ by a new variable $z$ as follows:
\begin{equation}
z = \frac{ 2 E R - {\mu}^2}{(2E)^{2/3}} \hspace{0.5cm}.
\end{equation}

\noindent 
Then Eq. (\ref{es6}) becomes
\begin{equation}
\frac{d^2 \Psi_{R}(z)}{dz^2} - z \hspace{0.12cm}\Psi_{R}(z) = 0\hspace{0.5cm},
\end{equation}

\noindent 
which is  Airy's differential equation \cite{Magnus}. This equation has a physically unacceptable solution  $\, \mbox{Bi}(z)\,$ that grows exponentially as $\, z\rightarrow \infty\,$. Thus, we pick the exponentially decreasing solution
\begin{equation}
\label{psiAiry}
\Psi_{R}(R)= \mbox{Ai}\Biggl( \frac{2ER-\mu^2}{(2E)^{2/3}}\Biggr)  \hspace{0.3cm}.
\end{equation}

%We shall discuss only the boundary condition (\ref{boundary1}). The %requirement $\, \psi_{R}(0)=0\,$ leads to

We shall restrict ourselves to the boundary condition (\ref{boundary1}), which leads to
\begin{equation}
\mbox{Ai}\Biggl(- \frac{\mu^2}{(2E)^{2/3}}\Biggr) = 0  \hspace{0.3cm}.
\end{equation}

\noindent 
Airy's function $\, \mbox{Ai}(z)\,$ has infinitely many negative zeros $\, z_n = -a_n\,$, with $\,a_n>0\,$, so that the allowed values of $\, E\, $ are
\begin{equation}
E_n = \frac{\mu^3}{2{a_n}^{3/2}}
 \hspace{0.3cm}.
\end{equation}

Thus, a complete  set of  solutions to equation (\ref{es2}) in the case $\, k=0\,$ is  
\begin{equation}
\label{psik=0}
\Psi_{n\,\mu}(R, \chi ,t)= \mbox{Ai} \Bigg(\frac{\mu\, R}{a_n^{1/2}} - a_n \Bigg)\, e^{i\,\mu\,\chi} \, e^{-i\, \mu^3\, t/2\, a_n^{3/2}}  \hspace{0.3cm}.
\end{equation}
 
\noindent 
By superposing such solutions one can construct finite-norm solutions to the Wheeler-DeWitt equation (\ref{es2}).

It is important to mention that Airy's function $ \mbox{Ai}(z)$ exhibits an oscillatory behaviour for $z<0$ ($R<\frac{a_n^{3/2}}{\mu}$) whereas for $z>0$ ($R>\frac{a_n^{3/2}}{\mu}$) it does not oscillate, but decreases monotonically  and for large $\, z\,$ becomes an exponentially damped function. This is displayed in Fig. 1. Thus, contrary to what is usually expected, the solutions (\ref{psik=0})
display a classical behaviour for small $\, R\,$ and a quantum behaviour for large $\, R\,$. The possibility of  existence of detectable quantum  gravitational effects in large universes is very remarkable, and this kind of phenomenon has been encountered recently in models arising from Kaluza-Klein theories \cite{Nelson}. From Eqs.(\ref{metric}) and (\ref{escala}) we find that $\, R(\tau ) \rightarrow \tau^{2/3}\,$ or $\, R(\tau ) \rightarrow \tau^{1/2}\,$ for large cosmic time $\, \tau\,$, depending on whether $\, p_{\epsilon}\neq 0\,$ or  $\, p_{\epsilon}= 0\,$ . 
Thus, the peculiar behaviour of the solutions (\ref{psik=0}) to the Wheeler-DeWitt equation takes place even in the case of an asymptotic 
power-law expansion $\, R(\tau ) \propto \tau^p\,$ with $\, p> 1/3\,$, contrary to what is suggested in \cite{Vink}. Therefore, the results according to which quantum gravitational effects are suppressed for large $\, R\,$ if $\, p>1/3\,$ seem to be a particularity of the minisuperspace considered in \cite{Vink}, and do not carry over to all minisuperspace models.

The above conclusion that the wave functions (\ref{psik=0}) exhibit a classical behaviour for small $\, R\,$ and a quantum behaviour for large 
$\, R\,$  is corroborated by a Bohmian analysis \cite{Nelson}. The classical potential $\, U = -\sqrt{h}\, {^{(3)}\!R}\,$ vanishes because the spacelike sections have no curvature for $\, k=0\,$. In its turn, the quantum potential associated with the wave functions (\ref{psik=0}) is
\begin{equation}
\label{potquantico}
Q(R)= -\frac{1}{2}\frac{1}{Ai(z)}\frac{d^2Ai(z)}{dR^2}= -\frac{\mu^2}{2 a_n}\Bigl(
\frac{\mu R}{a_n^{1/2}}-a_n \Bigr)
 \hspace{0.3cm},
\end{equation}

\noindent 
which is small for small $\, R\,$ and large for large $\, R\,$. Moreover, since Eq.(\ref{psik=0}) is of the form $\, Pe^{iS}\,$ with $\, S\,$ independent of $\, R\,$, the quantum Bohmian trajectory  defined by 
\begin{equation}
\label{Bohm}
p_R = \frac{\partial S}{\partial R}\,\,\,\,\, , \,\,\,\,\,
p_{\chi} = \frac{\partial S}{\partial \chi}
 \hspace{0.3cm},
\end{equation}

\vspace{0.3cm}

\noindent
is given by
\begin{equation}
\label{trajetoriaq}
R=C_1 \,\,\,\,\, , \,\,\,\,\,
\chi = \mu t + C_2
 \hspace{0.3cm},
\end{equation}

\vspace{0.3cm}

\noindent
where $\, C_1\,$ e $\, C_2\,$ are constants, and we have used $\, p_R = -{\dot R}\,$ and $\, p_{\chi} = {\dot \chi}\,$. A comparison with Eqs.(\ref{escala})-(\ref{vinculo1}) shows that $\, B\,$ and $\, p_{\epsilon}\,$ cannot both vanish for a nonvanishing coefficient $\, D=\mu/\sqrt{12}\,$. Therefore, Eq.(\ref{escala}) shows that the quantum trajectory differs from the classical one. The departure of the quantum trajectory from the classical one is more pronounced for large $\, t\,$, which means large $\, R\,$. Therefore, quantum effects are expected only for large values of the scale factor.

\section{FRW Model With Conformal Scalar Field and Dynamical Vacuum}

We now turn our attention to a FRW model filled with a  conformal scalar field and a vacuum fluid. Once again Schutz$^{,}$s variables and a FRW metric are used. As before, the ADM reduction takes the total action to the form \cite{Lemos2,Peleg}
\begin{equation}
S=\int_{}^{} dt (p_{a}\dot{a} + p_{\Phi}\dot{\Phi} + p_{\epsilon}\dot{\epsilon} +  p_{S}\dot{S} - N\cal{H})\hspace{0.5cm},
\end{equation}

\noindent 
where the super-Hamiltonian $\cal{H}$ is now given by
\begin{equation}
\label{sh}
{\cal{H}}=-\frac{{p_{a}}^2}{24} - 6ka^2 + \bigg(p_{\epsilon}^{\gamma}a^{-(3\gamma - 4)}\hspace{0.12cm}e^{S}\bigg) + \bigg(\frac{p_{\Phi}^2}{24} + 6k{\Phi}^2\bigg)\hspace{0.5cm}.
\end{equation}

If we take the perfect fluid as a vacuum fluid ($\gamma = 0$) and then make a canonical transformation to  new variables $S,T$ such that $T=-e^{-S}p_{S},\hspace{0.3cm} p_{T}=e^{S}$, the super-Hamiltonian  reduces to
\begin{equation}
{\cal{H}}= - \bigg(\frac{p_{a}^2}{24} + 6ka^2\bigg) + \bigg(\frac{p_{\Phi}^2}{24} +6k{\Phi}^2\bigg) + p_{T}a^4\hspace{0.5cm}.
\end{equation}

Upon defining as quantum operators
\begin{equation}
p_{a}\rightarrow -i\frac{\partial}{\partial a}\hspace{0.4cm},\hspace{0.4cm}p_{\Phi}\rightarrow -i\frac{\partial}{\partial \Phi}\hspace{0.4cm}\mbox{and}\hspace{0.4cm}p_{T}\rightarrow -i\frac{\partial}{\partial T}\hspace{0.5cm},
\end{equation}

\noindent 
we obtain the  Wheeler-DeWitt equation

\begin{displaymath}
\bigg(\frac{1}{24 a^4}\frac{{\partial}^2}{\partial a^2} - 6\frac{k}{a^2}\bigg)\Psi(a,\Phi,T) - \bigg(\frac{1}{24 a^4}\frac{{\partial}^2}{\partial {\Phi}^2} - 6k {\Phi}^2\bigg)\Psi(a,\Phi,T)
\end{displaymath}
\begin{equation}
\label{s1}
 = i\frac{\partial}{\partial T} \Psi(a,\Phi,T)\hspace{0.3cm}
\end{equation}

\noindent 
with $T$ playing the role of time.

The reparametrization 

\begin{equation}
a=\frac{R}{\sqrt{12}}\hspace{0.4cm},\hspace{0.4cm}\Phi=\frac{\chi}{\sqrt{12}}\hspace{0.4cm}\mbox{and}\hspace{0.4cm}T=\frac{t}{144}\hspace{0.5cm},
\end{equation}

\noindent 
takes  equation (\ref{s1}) to the form

\begin{displaymath}
\bigg(\frac{1}{2}\frac{{\partial}^2}{\partial R^2} - \frac{1}{2}k R^2\bigg)\Psi(R,\chi,t) - \bigg(\frac{1}{2}\frac{{\partial}^2}{\partial {\chi}^2} - \frac{1}{2}k{\chi}^2\bigg)\Psi(R,\chi,t)
\end{displaymath}
\begin{equation}
\label{s2}
 = R^4\hspace{0.2cm}i\hspace{0.15cm}\frac{\partial}{\partial t} \Psi(R,\chi,t)\hspace{0.3cm}.
\end{equation}

Once more we try to solve (\ref{s2}) by separation of variables by writing $\Psi(R,\chi,t)=\Psi_{R}(R)\Psi_{\chi}(\chi)e^{-iEt}$. This gives rise to the  ordinary differential equations
\begin{equation}
\label{s3}
\frac{1}{2\Psi_{R}(R)}\frac{{\partial}^2 \Psi_{R}(R)}{\partial R^2} -\frac{1}{2}kR^2 - ER^4 = -\lambda \hspace{0.3cm},
\end{equation}

\begin{equation}
\label{s4}
\frac{1}{2\Psi_{\chi}(\chi)}\frac{{\partial}^2 \Psi_{\chi}(\chi)}{\partial {\chi}^2} -\frac{1}{2}k{\chi}^2  = -\lambda \hspace{0.3cm}.
\end{equation}

\noindent 
As pointed out in \cite{Lemos2}, the parameter $\, E\,$ plays the role of the cosmological constant $\, \Lambda\,$.

We consider only the case of closed universes, that is $k=1$. In the special case $E=0$  equations (\ref{s3}) and (\ref{s4})  are equivalent to 
\begin{equation}
\frac{1}{2}\bigg( {\chi}^2 - {p_{\chi}}^2 - {R}^2 - {p_{R}}^2 \bigg)\Psi(R,\chi) = 0
\end{equation}

\noindent 
or
\begin{equation}
\bigg({H_{\chi}}^{\mbox{oscil}} - {H_{R}}^{\mbox{oscil}}\bigg)\hspace{0.3cm}\Psi(R,\chi) = 0 \hspace{0.3cm}.
\end{equation}

Apart from the restriction $R>0$, the above Hamiltonian is exactly the difference of the Hamiltonians of two harmonic oscillators. Thus, we obtain again the Hawking-Page wormhole solutions 
\begin{equation}
\label{wormhole}
 \Psi_{n}(R,\chi)=\varphi_{n}(R)\hspace{0.2cm}\varphi_{n}(\chi)
\end{equation}

\noindent 
for the wave function of the Universe, where $\, n\,$ is odd or even according to which of the boundary conditions (\ref{boundary1}) or 
(\ref{boundary2}) is chosen. The wave functions  (\ref{wormhole})  are 
normalizable in the inner product
\begin{equation}
\label{inner2}
(\Psi ,\Phi ) = \int_{-\infty}^{\infty}d\chi \int_0^{\infty} dR \, R^4\, \Psi^*(R,\chi ,t )\, \Phi (R,\chi ,t)\hspace{0.3cm}.
\end{equation}
\newline

\noindent
This inner product turns the Wheeler-DeWitt equation (\ref{s2}) into a bona fide Schr\"{o}dinger equation with a self-adjoint hamiltonian operator, so that $\, t\,$ can be interpreted as a genuine time \cite{Lemos2}.

No analytical solutions are known for the general case
$\, E\neq 0\,$, in which Eq.(\ref{s3}) becomes
\begin{equation}
{\hat{H}}\Psi_{R}(R)\equiv\bigg(\frac{{{\hat{p}}_{R}}^2}{2} + V(R)\bigg)\Psi_{R}(R)=\lambda \Psi_{R}(R)\hspace{0.5cm},
\label{anarmonico}
\end{equation}

\noindent 
where $\, \lambda = n+1/2\,$ and 
\begin{equation}
\label{potential}
V(R) = \frac{1}{2}kR^2 + ER^4
\end{equation}

\noindent 
is a quartic anharmonic oscillator potential. Equation (\ref{anarmonico}) is a rather curious type of eigenvalue problem. Since $\,\lambda\,$ is given, the ``coupling constant" $\, E\, $ that appears in the quartic potential must be  adjusted to values $\, E_n\, $ such that  the eigenvalues of the Hamiltonian operator are exactly those of the usual harmonic oscillator, namely $\, \lambda = n+1/2\,$. 

Numerical solutions to many eigenvalues of the quartic anharmonic oscillator are  known with high precision \cite{art,Bacus}. Since the potential (\ref{potential}) is an even function of $\, R\,$, it is known \cite{Schiff} that  the eigenfunctions of $\,{\hat{H}} \,$ may be separated in even or odd solutions, so that one of the boundary conditions $\, {\psi}(0)=0\,$ or $\, {\psi}^{\prime}(0)=0\,$ can be satisfied for any allowed value of $\, E\,$.

From the Hellmann-Feynman theorem \cite{Merzbacher} it follows immediately that $\, \lambda\,$ is a monotonically increasing function of $\, E\,$, and it is clear physically that $\, \lambda \rightarrow \infty\,$ as 
$\, E \rightarrow \infty\,$. Thus, the spectrum of the cosmological constant $\, E\,$ is discrete, and contains only positive eigenvalues, as one infers from the semiquantitative Fig. 2, which was obtained by interpolating the data extracted from  Table I of \cite{art}. The allowed values of $E$ are the abscissas of the points at which the dashed horizontal lines $\lambda =n+1/2$ intersect the curves $\lambda_m(E)$.
The apparent  degree os degeneracy is $\, n+1\,$ for the $n$-th eigenvalue $\, E_n\,$, but the true degree of degeneracy is reduced by the self-adjointness boundary conditions.

Therefore, there exists a denumerable set of normalizable (``physical") quantum states with  positive values of the cosmological constant which are endowed  with all the properties of quantum wormhole states. This is made possible by the presence of the conformally coupled scalar field, since in pure de Sitter's model normalizable states with a sharp value of the cosmological constant are forbidden \cite{Lemos2}.

\section{Summary and Final Comments}

We have quantized a FRW universe  filled with dust and a  conformally invariant scalar field. For the closed model  a discrete set of wormhole quantum states has been found. Such states are described by wave functions with finite norm with respect to an inner product defined by integration over a configuration space that does not include the time variable. For  flat spacelike sections, solutions to the Wheeler-DeWitt equation have been discovered which, contrary to our intuitive expectations, present quantum  behaviour for large  values of the scale factor and behave classically for small values of the scale factor. This has been confirmed by an analysis in the spirit of Bohm's causal interpretation of quantum mechanics, and disagrees with suggestions according to which quantum gravitational effects are suppressed for large $\, R\,$ in the case of an asymptotic power law expansion $\, R(\tau ) \propto \tau^p\,$ with $\, p> 1/3\,$. 

We have also quantized a  FRW model with a conformally invariant scalar field and a nonvanishing cosmological constant  dynamically introduced by regarding the vacuum  as a perfect fluid with equation of state $p = - \rho$. In the case of a closed universe we have obtained finite-norm solutions to the  Wheeler-DeWitt equation with definite values of the cosmological constant that represent wormholes. In contrast to  the plain de Sitter model, there exist realizable quantum states for certain discrete positive values of the cosmological constant.

Our study has indicated that wormhole states appear to be a common occurrence in quantum cosmological models with a very simple matter content. Moreover, no exotic material sources are necessary to give rise to expanding quantum cosmological models showing quantum effects only for large values of the scale factor.

\vskip 1cm

\noindent 
ACKNOWLEDGEMENT

\noindent 
This work was partially supported by Conselho Nacional de Desenvolvimento Cient\'{\i}fico e Tecnol\'ogico, CNPq, Brazil.

\newpage

{\huge{\bf{Caption Lists}}}
 
\vskip 2.5cm

Figure 1: Plot of the radial part of ${{\Psi}_{n\mu}}$ for ${{\mu}=1}$ and ${n=4}$, displaying oscillatory behaviour for small values of the scalefactor and exponential damping for large values of the scale factor.

\vspace{1.5cm}

Figure 2: Plot of the first few eigenvalues ${\lambda}_n$ of equation (68) showing dependence on ``coupling constant" $E$.

\end{document}